# Rate Constrained Random Access over a Fading Channel

Nitin Salodkar and Abhay Karandikar, *Member, IEEE*

*Abstract*—In this paper, we consider uplink transmissions involving multiple users communicating with a base station over a fading channel. We assume that the base station does not coordinate the transmissions of the users and hence the users employ random access communication. The situation is modeled as a non-cooperative repeated game with incomplete information. Each user attempts to minimize its long term power consumption subject to a minimum rate requirement. We propose a two timescale stochastic gradient algorithm (TTSGA) for tuning the users' transmission probabilities. The algorithm includes a 'waterfilling threshold update mechanism' that ensures that the rate constraints are satisfied. We prove that under the algorithm, the users' transmission probabilities converge to a Nash equilibrium. Moreover, we also prove that the rate constraints are satisfied; this is also demonstrated using simulation studies.

## I. Introduction

Wireless networks have witnessed large scale proliferation in the recent years. Apart from voice applications, data applications such as World Wide Web (WWW), email etc. are also extremely popular with the users. Different applications have different Quality of Service (QoS) requirements from the network in order to perform satisfactorily. These QoS requirements can be in terms of parameters such as the delivered rate, delay or delay jitter. In this paper, we consider providing QoS (average rate) guarantees to web applications such as WWW operating over a single cell wireless network. This entails addressing the following important issues:

- Medium access control (MAC): Multiple users need to access the common wireless channel simultaneously in order to communicate with a common receiver such as a base station or an access point. The access mechanism must be so designed that it satisfies the QoS requirements of user applications.
- Challenges offered by the Wireless Medium: Wireless channel is characterized by decay of signal strength due to distance (path loss), obstructions due to objects such as buildings and hills (shadowing), and constructive and destructive interference caused by copies of the same signal received over multiple paths (multipath fading), possibly with time varying path lengths, resulting in a time varying channel condition [1]. These phenomena distort the signal in an unpredictable manner and can cause packet errors at the receiver. The time varying wireless channel poses significant challenges for design of efficient and reliable communication systems.

Various MAC protocols have been proposed in the literature that attempt to satisfy different application requirements based on their packet arrival characteristics and QoS attributes. These protocols can be classified into following two types:

- Fixed resource allocation protocols: These protocols assign fixed amount of resources to the users by means of orthogonal or near orthogonal channels. They require a central scheduling entity (like a base station) that performs the channel allocation task. Examples include Time Division Multiple Access (TDMA), Frequency Division Multiple Access (FDMA), Code Division Multiple Access (CDMA) and Orthogonal Frequency Division Multiple Access (OFDMA). These have been implemented in cellular systems [1].
- Random access protocols: In these protocols, users access the channel randomly. Users vary their channel access probabilities or access times based on limited feedback from the channel. Since the users take transmission decision based on local information available with them, these protocols are suited for distributed implementation. Random access protocols have been implemented in cellular systems, satellite communication systems and multitap bus among others. These have been well studied for the past several decades. [2], [3] provide excellent textbook treatment of random access protocols.

Recent research on designing efficient transmission strategies over fading wireless channels has revealed a lot of interesting insights which suggest that fading can be considered as an opportunity instead of it being treated as an adversary. Users located at diverse locations are likely to perceive diverse channel conditions (multiuser diversity). When the number of users is large, the probability that a certain user perceives very good channel condition is close to one [4]. This multiuser diversity has been exploited for designing efficient *cross layer* schemes [5] that utilize this information from the physical layer in order to make efficient scheduling decisions at the MAC/network layer (e.g., scheduling the user perceiving the 'best' channel condition in each slot). Several studies have demonstrated that this strategy substantially improves performance in terms of throughput [4], [6].

In this paper, our objective is to design a random access scheme for providing average rate guarantees to the contending users. We consider the uplink scenario where multiple users

Nitin Salodkar is currently with General Motors India Science Lab, ITPL, Bangalore India - 560066. Email: nitin.salodkar@gm.com Work done when Nitin was a doctoral candidate with Department of Computer Science and Engineering, IIT Bombay.

Abhay Karandikar is with Department of Electrical Engineering, IIT Bombay, Powai, Mumbai India - 400076. Email: karandi@ee.iitb.ac.in Work supported in part by Tata Teleservices-IIT Bombay Center for Excellence in Telecom.

transmit to a single base station over the wireless channel. The users take advantage of the opportunities provided by the wireless channel by varying their transmission powers based on their channel condition ('channel aware' transmission). Since wireless devices are battery powered with limited battery energy, in order to conserve energy, the objective is to devise a random access and power control scheme wherein the long term average power consumption for each user is minimized subject to satisfying the average rate requirement of each user.

There have been several attempts to design channel aware random access schemes in order to improve the performance of these schemes. Existing work on channel aware random access schemes can be classified into the following classes:

- Signal processing and diversity techniques to correctly decode received packets as in [7], [8], [9], [10].
- Work that advocates the adaption of retransmission probabilities of users either through 'Splitting algorithms' that adapt the set of contending users based on feedback from the channel as in [11], [12] or through 'channel aware ALOHA' schemes as in [13], [14], [15], [16].

Our work falls under the latter category.

### A. Related Work

In this section we review representative research that advocates adapting retransmission probabilities of users under channel aware ALOHA schemes.

In [13], [17], the authors attempt to exploit multiuser diversity in a distributed fashion with only local channel information, i.e., each user is aware of its own channel condition only. The authors propose a channel aware ALOHA protocol and provide a throughput analysis of the proposed protocol under an infinitely backlogged model. In [15], the authors consider symmetric as well as asymmetric fading. The authors propose a *binary* scheduling algorithm where users access the channel when the corresponding channel condition is above a certain threshold and prove that it maximizes sum throughput under symmetric fading. Moreover, for asymmetric fading, they prove that binary scheduling maximizes the sum of log of average throughput of the users and is fair in the long run. Furthermore, they also consider channels with memory and provide simple extensions of the binary scheduling algorithm. In [14], the authors study distributed schemes for exploiting multiuser diversity in the uplink (multipoint to point) context. They propose a channel aware ALOHA scheme where the transmission probability is a function of channel state information. They characterize the maximum stable throughput for such a system with both finite as well as infinite user models. In [16], the author defines an interference-dominating wireless network as the one in which a receiver could simultaneously receive a number of packets from a variety of transmitters, as long as the signal-to-interference-plus-noise ratio exceeds a predetermined threshold. The author proposes an analytical approach to derive the exact value of saturation throughput of slotted ALOHA in such an interference-dominating wireless ad-hoc network.

In addition to the above, there has been a lot of interest in power control techniques over random access wireless networks. Game theory [18] serves as a useful tool for designing these power control schemes. Moreover, it also provides a useful framework for designing access control protocols with provision for information exchange (cooperative game) as well as no information exchange (non-cooperative game) between users. Game theoretic models have been extensively applied [19], [20], [21], [22]. The slotted ALOHA protocol has been modeled both as a non-cooperative game as well as a cooperative game. Various objectives like delay minimization [19], throughput maximization [19], [20], [21] have been considered. Power control coupled with retransmission control has been variously studied in [23], [24], [25]. In [26], the authors analyze the equilibrium points achieved by a non-cooperative group of users that have a certain QoS requirement and willingness to pay. The reader is referred to [24], [27], [28] for further information on applications of game theory for modeling the random access problem.

The model considered in this paper is similar to that in [17]. However, in [17], the authors consider the throughput scaling under long term and short term power constraints, while we consider a different problem of minimizing the long term power expenditure of each user subject to satisfying the rate constraint of each user. The problem formulated in this paper is similar to that of a recent work in [29]. In [29] the authors assume that users transmit at constant power, while our formulation also includes power control. Moreover, while in [29], the authors' emphasis is on *analysis* of the Nash Equilibria of the game involving a group of non-cooperative users sharing a channel and desiring certain long term throughput, we adopt a *prescriptive* approach. We focus on algorithm design and show that the Two Timescale Stochastic Gradient Algorithm (TTSGA) suggested by us converges to the Nash equilibrium of the game under certain conditions.

### B. Our Contributions

In this paper, our objective is to provide QoS (average rate) guarantees to users while taking advantage of the opportunities provided by the fading wireless channel. We assume that the distribution of the channel fading process is not known to the users. We assume that each user accesses the channel independently of others based on its Channel State Information (CSI) only. Moreover, the users are not aware of the rate requirements and the channel conditions of the other users. Furthermore, there is no mechanism for information exchange between the users. This situation is modeled as a constrained repeated non-cooperative game, where each user has an objective of minimizing the long term average power expenditure subject to achieving a certain long term average rate. The uses modulate their transmission rates based on the CSI fed by the base station. We propose an iterative primal-dual technique for tuning the transmission probabilities of the users and ensuring that the constraints are satisfied. The primal variable is the transmission probability while the dual is the waterfilling threshold that adjusts the average transmission power for ensuring that the rate constraint is satisfied. Our contributions can be summarized as follows:

- We formulate the user problem where the objective is to minimize average power consumption subject to an

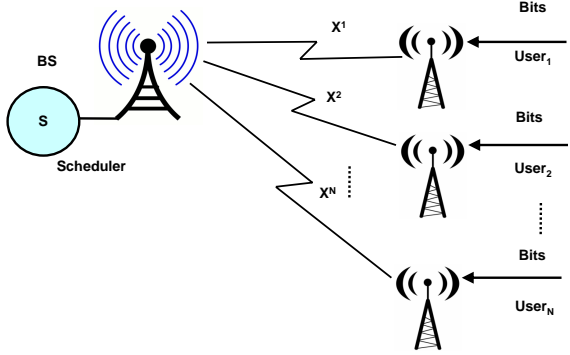

Fig. 1. Uplink transmission scenario

- average rate constraint as a constrained repeated non-cooperative game where each user has knowledge of its rate requirement and CSI only.
- We propose a Two Timescale Stochastic Gradient Algorithm (TTSGA) for iteratively tuning the transmission probability. This algorithm iterates the transmission probability in the direction of the gradient of the average power consumption. Moreover, the 'waterfilling threshold' is tuned on a faster timescale to ensure that the rate constraints are satisfied.
- We prove that under TTSGA, the transmission probability iterates converge to a Nash equilibrium provided certain conditions are met.
- The approach ensures that the rate constraints are met. This is proved analytically and validated through simulations.

The rest of the paper is organized as follows. Section II provides details regarding the system model. We formulate the problem as a constrained repeated game in Section III and then motivate a solution within the random access framework. In Section IV, we motivate the solution strategy and propose TTSGA. In Section VI we analyze certain properties of the algorithm and prove that under the algorithm, the transmission strategies of users converge to a Nash equilibrium. Moreover at equilibrium, the rate constraints are satisfied. We present the simulation results in Section VII. In Section VIII, we discuss implementation aspects of the algorithm within IEEE 802.11 framework. We conclude in Section IX.

## II. SYSTEM MODEL

We consider an uplink scenario similar to that in [17] as depicted in Figure 1 where $N$ users communicate with a base station. We consider a time slotted system, i.e., time is divided into slots of equal duration normalized to unity. There can be multiple flows between a user and the base station. However, for the sake of notational simplicity, we assume that only one flow exists between a user and the base station. The analysis can be easily extended to more general cases.

We assume that the system operates in a *distributed* fashion. Hence we assume that the base station does not coordinate the transmissions of the users. In each slot $n$, user $i$ transmits with a certain probability $\theta_n^i$ to the base station. If more than one user transmits in a slot, then all transmissions are unsuccessful, i.e., there is a *collision* at the base station. We assume that each user receives a $(0, 1, e)$ feedback in each slot, where $0$ denotes that there is no transmission in the slot, $1$ denotes successful transmission and $e$ denotes collision or unsuccessful transmission. We assume that this feedback is immediate and error free. In practice, this information can be conveyed through acknowledgement messages sent by the base station over a feedback channel.

We assume a wireless channel with block fading [30]. Under this model, if $\chi_n^i$ is the transmitted signal by user $i$ in slot $n$, then the signal $Y_n^i$ received by the base station in slot $n$ can be expressed as:

$$Y_n^i = H_n^i \chi_n^i + Z_n, \quad (1)$$

where $Z_n$ is complex Additive White Gaussian Noise (AWGN) at the base station. $H_n^i$ is the channel gain and we denote $X_n^i = |H_n^i|^2 \in \mathbb{X}$ as the channel state for user $i$ in slot $n$. We assume that the users possess perfect knowledge of channel state $X_n^i$ in each slot [1]. Moreover, the distribution of $X_n^i$ is not known to user $i$. We discuss possible mechanisms for conveying information such as successful reception of a packet and CSI in Section VIII.

All packets are assumed to be of equal size, say, $\ell$ bits. We assume a backlogged model, i.e., users always have packets to transmit. Let $U_n^i$ denote the number of packets that user $i$ can transmit *reliably* to the base station in slot $n$. Since the slot duration is normalized to unity, $U_n^i$ can be considered to be the transmission rate in slot $n$. In practice, this can be determined based on the modulation or coding scheme employed at the physical layer. Let $P^i(X_n^i, U_n^i)$ denote the power consumed by user $i$ while transmitting at rate $U_n^i$ when the channel state perceived by the base station is $X_n^i$.

## III. PROBLEM FORMULATION

In this section, we formulate the problem as a repeated non-cooperative game. We begin by considering the $i$th user. The long term power consumption for user $i$ can be expressed as:

$$\bar{\mathcal{P}}^i = \limsup_{M \to \infty} \frac{1}{M} \sum_{n=1}^{M} \theta_n^i P(X_n^i, U_n^i). \quad (2)$$

where $\theta_n^i$ denotes the transmission probability for user $i$ in slot $n$. Let $\beta_n^i$ denote the probability of successful transmission for user $i$ in slot $n$. The long term throughput or rate achieved by user $i$ can be expressed as:

$$\bar{\mathcal{U}}^i = \liminf_{M \to \infty} \frac{1}{M} \sum_{n=1}^{M} \beta_n^i U_n^i. \quad (3)$$

The optimization problem for user user $i$ can be expressed as:

$$\text{Minimize } \bar{\mathcal{P}}^i \text{ subject to } \bar{\mathcal{U}}^i \geq \bar{\rho}^i, \quad (4)$$

where $\bar{\rho}^i$ denotes the long term average rate requirement for user $i$.

---

[1] In practice, the users perform channel estimation for downlink transmissions using the pilot symbols transmitted by the base station. In a Time Division Duplex (TDD) system because of symmetry, these estimates can be used for uplink transmissions as well.

We are interested in determining an 'equilibrium' or 'steady state' transmission probability for each user (say $\theta^{i,*}$ for user $i$) and a channel state dependent power control policy such that if a user transmits at this probability with the power prescribed by the power control policy, the average power is minimized and the rate constraint is satisfied. Note that the above problem can be viewed as a rate constrained random access problem where each user $i$ solves a similar optimization problem.

*Remark 1:* Let $\mathcal{P}^i_{\theta^i}$ denote the average power expenditure for user $i$ when the transmission probability is fixed at $\theta^i$. It consists of two parts: 'useful' power, i.e., the power consumed in successful transmissions and 'wasted' power, i.e., the power wasted in collisions. Let $\mathcal{P}^i_{\beta^i}$ and $\mathcal{P}^i_{\varepsilon^i}$ denote the 'useful' power and 'wasted' power respectively when the success and collision probabilities are $\beta^i$ and $\varepsilon^i$ respectively. The problem for each user is to determine an equilibrium transmission probability $\theta^{i,*}$ such that $\mathcal{P}^i_{\theta^{i,*}} = \mathcal{P}^i_{\beta^{i,*}} + \mathcal{P}^i_{\varepsilon^{i,*}}$ is minimum.

The problems being solved by $N$ users are not independent. Transmission probability of one user (say user $i$) impacts the collision/success probabilities of all the other users. This affects their transmission probability which in turn impacts the transmission probability of user $i$. Thus, this is a game situation [18]. Each user attempts to minimize its own disutility (power) subject to its rate requirement. We assume that a user is not aware of the rate requirement of the other users. Moreover, since there is no provision for information exchange between the users, the CSI is also localized at the users, i.e., a user is not aware of the CSI of other users. Furthermore, since the users are only provided with a $(0,1,e)$ feedback by the base station, the users cannot fully observe the actions taken by the other users. We view the situation as a repeated non-cooperative game with incomplete information. The solution concept that we target is that of the Nash equilibrium. In this case, at equilibrium, each user's transmission strategy in the long run can be viewed as a 'best response' to the long term transmission strategies of the rest of the users.

## IV. SINGLE USER SCENARIO

Before proceeding with the multiuser scenario, we analyze the single user scenario that provides us with key insights that aid in designing an efficient multiuser solution.

### A. Single User Scenario

Consider a single user scenario as depicted in Figure 2. We assume that the user user is split into two virtual entities: a transmitter and a scheduler. The user transmits over a block fading channel in a time slotted system. The scheduler has an objective of minimizing the long term average power expenditure subject to an average rate constraint (say $\bar{\rho}$)[2]. To meet this objective, in each slot, the scheduler determines the transmission rate (say $U_n$) based on the channel state (say $X_n$) and directs the transmitter to transmit at that rate. However, with a certain probability $\beta$, the transmitter is unable to proceed with the transmission. In this case, the long term throughput achieved by the user can be expressed as:

$$\bar{U} = \liminf_{M \to \infty} \frac{1}{M} \sum_{n=1}^{M} \beta U_n. \quad (5)$$

[2]Since we deal with the single user scenario, in this section, we omit the superscript from the notation for notational simplicity.

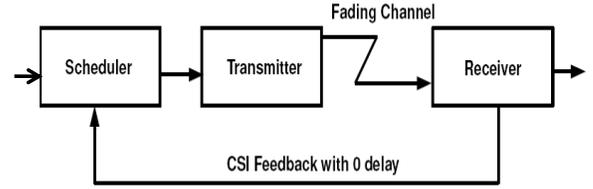

Fig. 2. Single user scenario

Moreover, the long term power consumption can be expressed as:

$$\bar{P} = \limsup_{M \to \infty} \frac{1}{M} \sum_{n=1}^{M} \beta P(X_n, U_n).$$

The problem can be precisely expressed as:

$$\text{Minimize } \bar{\mathcal{P}} \text{ subject to } \bar{\mathcal{U}} \geq \bar{\rho}. \quad (6)$$

Note that the present problem is a generalization of dual of the problem considered in [31]. The analysis for determining the optimal transmission policy proceeds on similar lines as in [31]. The constrained problem in (6) can be converted into an unconstrained problem using the Lagrangian approach [32]. The unconstrained problem can be expressed as:

$$\text{Minimize } \bar{\mathcal{P}} - \lambda(\bar{\mathcal{U}} - \bar{\rho}), \quad (7)$$

where $\lambda \geq 0$ is referred to as the Lagrange Multiplier (LM).

It can be verified that for the present problem, the optimal power allocation for a channel state $X_n = x$ can be expressed as:

$$P_w(x) = (\lambda^* - \frac{N_0}{x}), \quad (8)$$

where $N_0$ is the power spectral density of the AWGN at the base station. The rate $U_n$ under channel state $X_n$ can then be determined based on this power allocation. Note that $P_w(\cdot)$ depends on the optimal LM $\lambda^*$ which in turn is a function of the transmission probability $\beta$. This optimal LM $\lambda^*$, which we also refer to as the 'waterfilling threshold', however, needs to be determined.

*Remark 2:* Note that in the single user scenario, the transmitter transmits with a probability $\beta$ and remains idle with probability $1-\beta$. The slots where the transmitter does not transmit can be considered to be equivalent to the slots where the channel state is extremely poor. For such slots, the Goldsmith-Varaiya (G-V) waterfilling power allocation scheme [31] does not transmit. It is easy to see that the G-V scheme is optimal for this case also, albeit with an appropriately scaled LM or the 'waterfilling threshold' as compared to the case where the transmitter transmits in each slot. Effectively, as $\beta$ decreases, the scheduler perceives a progressively poorer channel. This results in a correspondingly higher value for the optimal LM $\lambda^*$ resulting in higher power allocation in each channel state for satisfying the rate constraint.

*Remark 3:* It can be argued that for the single user scenario, in the optimal solution, the constraint is met with equality. We know that power is an increasing convex function of the transmission rate. Now, suppose that the constraint is not met with equality, then a further reduction in power consumption can be achieved by transmitting at lower rates and then meeting the constraint with equality. This implies that if the constraint is not met with equality, the solution is not optimal. Hence, in the optimal solution, the constraint is met with equality.

*Remark 4:* Note that for the single user problem considered in (6), the waterfilling power allocation scheme (8) is optimal. For different values of transmission probability, the optimal LM $\lambda^*$ takes different values. Hence in the multiuser situation, we employ (8) for each user user $i$ for determining the transmission power at a given channel state. The task that remains is to determine the equilibrium transmission probability $\theta^{i,*}$ (and thereby the success probability $\beta^{i,*}$) and the corresponding constraint satisfying LM $\lambda^{i,*}$ for each user.

## V. Iterative Approach for Determining the Equilibrium Transmission Strategy in the Multiuser Scenario

In the previous section, we discussed the single user scenario. In this section, we focus our attention back to the multiuser scenario. In determining the equilibrium transmission probability within the multiuser setting, a user attempts to address the following tradeoff: if it were to transmit with too high probability, there could be too many collisions and wastage of power, on the other hand if it were to transmit with too low probability, it would amount to transmitting at higher power in each channel state in order to satisfy the rate constraint. The users attempt to achieve a balance between these conflicting objectives for arriving at a solution.

We now suggest a strategy for determining the equilibrium transmission probabilities. The essence of the solution strategy is the following: each user tunes its transmission probability iteratively so as to arrive at a Nash equilibrium. The users adapt their transmission probabilities in the direction of the gradient of average power expenditure. Moreover, the waterfilling threshold is also iteratively tuned so as to ensure that the rate constraints are satisfied. Before providing details of the solution strategy, we convert the constrained problem in (4) into an unconstrained problem using the Lagrangian approach [32].

### A. Lagrangian Approach

Let $\lambda^i$ be a Lagrange Multiplier (LM). The unconstrained problem (corresponding to the problem in (4)) can be stated as:

$$\text{Minimize } \mathcal{L}^i(\theta^i, \lambda^i) = \bar{P}^i - \lambda^i(\bar{U}^i - \bar{\rho}^i). \quad (9)$$

The objective is to determine the saddle point of the Lagrangian $\mathcal{L}^i(\theta^i, \lambda^i)$, i.e., to determine $\theta^{i,*}$ and $\lambda^{i,*}$ such that the following saddle point optimality conditions are satisfied:

$$\mathcal{L}^i(\theta^{i,*}, \lambda^i) \geq \mathcal{L}^i(\theta^{i,*}, \lambda^{i,*}) \geq \mathcal{L}^i(\theta^i, \lambda^{i,*}). \quad (10)$$

Note that the LM also acts as the waterfilling threshold. Hence, determining the equilibrium LM also results in determining the equilibrium waterfilling power allocation.

### B. Two Timescale Stochastic Gradient Algorithm (TTSGA)

We now present a stochastic gradient algorithm for the problem in (4) for a given user $i$. The essence of the algorithm is the following: in each slot a user $i$ determines its transmission power $P_w^i(X_n^i)$ based on its channel state $X_n^i$ using (8). User $i$ transmits with probability $\theta_n^i$ that serves as an estimate of the equilibrium transmission probability $\theta^{i,*}$. User $i$ tunes this estimate in the direction of the gradient of the Lagrangian in (9). In Section VI we prove that this algorithm leads to a Nash equilibrium.

The Lagrangian in (9) can be expressed as:

$$\begin{aligned}
\mathcal{L}^i(\theta^i, \lambda^i) &= \limsup_{M \to \infty} \frac{1}{M} \sum_{n=1}^{M} \left( \theta_n^i P_w(X_n^i) - \lambda^i(\beta_n^i U_n^i - \bar{\rho}^i) \right) \\
&= \limsup_{M \to \infty} \frac{1}{M} \sum_{n=1}^{M} g(X_n^i, U_n^i, \theta_n^i, \beta_n^i, \lambda_n^i), \quad (11)
\end{aligned}$$

where we refer to $g(\cdot, \cdot, \cdot, \cdot, \cdot)$ as the immediate cost function. Let $\nabla^{\theta^i} L^i(\cdot, \cdot)$ and $\nabla^{\lambda^i} L^i(\cdot, \cdot)$ denote the partial gradient of $L^i(\cdot, \cdot)$ w.r.t $\theta^i$ and $\lambda^i$ respectively. At the saddle point, $\theta^{i,*}$ and $\lambda^{i,*}$ satisfy the following conditions:

$$\nabla^{\theta^i} \mathcal{L}^i(\theta^i, \lambda^i)\Big|_{\theta^i = \theta^{i,*}} = 0, \quad (12)$$

$$\nabla^{\lambda^i} \mathcal{L}^i(\theta^i, \lambda)\Big|_{\lambda^i = \lambda^{i,*}} = 0, \quad (13)$$

and the complementary slackness condition,

$$\lambda^{i,*}(\bar{\mathcal{U}}^i - \bar{\rho}^i) = 0. \quad (14)$$

Note that the Lagrangian in (11) is a time average of the immediate cost function that can not be determined a priori in a real time implementation setup. If the equilibrium LM $\lambda^{i,*}$ is known, we can use an iterative method that improves its estimate of *only* the equilibrium $\theta^i$. Since the equilibrium LM $\lambda^{i,*}$ is also not known, we resort to a primal-dual method that determines *both* $\theta^{i,*}$ and $\lambda^{i,*}$ iteratively [32]. In order to ensure convergence of $\theta^i$ and $\lambda^i$ iterates to the equilibrium $\theta^{i,*}$ and $\lambda^{i,*}$, the iterations proceed at different *timescales*, i.e., the $\theta^i$ and $\lambda^i$ values are updated at different rates [33]. We iterate $\theta^i$ on a slower timescale, and $\lambda^i$ on a faster timescale. This implies that $\theta^i$ is maintained constant for a large number of $\lambda^i$ iterations. More specifically, as viewed from the $\lambda^i$ iteration, the $\theta^i$ iterates appear to be almost constant while as viewed from the $\theta^i$ iteration, the $\lambda^i$ iterates appear to be converged to the optimal value for the current value of $\theta^i$. The can be done in two ways. The first way is to physically separate the timescales by updating the LM in each time slot and updating the probability after a large number of time slots. The second way is by updating both quantities in each time slot, but carefully selecting the update sequences employed in the iteration. It can be shown that this has the same effect as that of the physical separation of time scales [33]. Let $\{a_n\}$ and $\{c_n\}$ be two positive sequences that have the following properties:

$$a_n \to 0, c_n \to 0; \quad \sum_n a_n = \infty, \quad \sum_n c_n = \infty;$$
$$\sum_k (a_n)^2 < \infty; \quad \sum_n (c_n)^2 < \infty. \quad (15)$$

Fix $\theta^i = \theta$. The equilibrium LM for this transmission probability that ensures that the complementary slackness condition (14) is satisfied can be determined using the following iteration carried out in each slot:

$$\lambda^i_{n+1} = \lambda^i_n - a_n(J^i_n U^i_n - \bar{\rho}^i), \ \lambda^i_n \geq 0 \ \forall n, \quad (16)$$

where $J^i_n$ is an indicator variable that is set to 1 if user $i$ transmission is successful in slot $n$. (16) forms the 'waterfilling threshold update mechanism'. $\lambda^i$ is the waterfilling threshold for user $i$, increasing this threshold results in an increase in power consumption, while decreasing this threshold results in a decrease in the average power consumption.

We now describe an approach for iterating the transmission probability $\theta^i$ in the direction of the gradient of the average power expenditure. This approach also involves separating the $\theta^i$ and $\lambda^i$ update timescales by carefully selecting the update sequences. It involves transmitting with probability $\theta^i + \delta$ in a odd numbered slots and with probability $\theta^i - \delta$ in even numbered slots at powers recommended by the waterfilling power allocation scheme (8). Let $(2n-1)$ and $(2n)$ refer to odd and even numbered slots respectively. Hence a user $i$ transmits with probability $(\theta^i_{2n-1}+\delta), \delta > 0, \delta << 1$, and $(\theta^i_{2n-1}-\delta)$ in odd and even numbered slots respectively. Based on the finite differences method [34], the gradient of the average power expenditure is determined and the transmission probability is updated in the direction of this gradient in odd numbered slots. This update equation can be expressed as:

$$\theta^i_{2n+1} = \pi_1 \left[ \theta^i_{2n-1} - c_{2n-1} \left( \frac{\mathcal{P}^i_{2n-1} - \mathcal{P}^i_{2n}}{2\delta} \right) \right], \quad (17)$$

where $\mathcal{P}^i_n$ is an estimate of the average power expenditure for user $i$ in slot $n$. Note that we do not have access to this estimate of the average power expenditure. In order to address this issue, we carry out simultaneous averaging of the power consumed. $\mathcal{P}^i_n$ is thus computed using the following recursive equation:

$$\mathcal{P}^i_{n+1} = \mathcal{P}^i_n + b_n \theta^i_n P^i_w(X^i_n), \quad (18)$$

where $b_n$ is an update sequence that has the same properties as those of $a_n$ and $c_n$ in (15).

We update $\mathcal{P}^i$ on a faster timescale as compared to the probability update timescale. This is done by imposing additional properties on update sequences $b_n$ and $c_n$ explained below. This ensures that as viewed from the average power update timescale, the transmission probability appears to be almost constant; the physical interpretation being that one is computing the average power expenditure for a certain large time interval with fixed value of the transmission probability. Note that in (18), $\theta^i$ can be considered to be the current quasi-static value of the transmission probability as seen from the power update timescale. LM is updated at the fastest timescale because it determines the waterfilling threshold. This guarantees that the scheme uses the correct waterfilling threshold and hence the transmission power $P^i_w(\cdot)$ is the correct power which in turn leads to correct average power values $\mathcal{P}^i$.

The different timescales specified above can be realized by imposing the following additional requirements on the update sequences $\{a_n\}, \{b_n\}, \{c_n\}$ [33]:

$$\frac{b_n}{a_n} \to 0; \quad \frac{c_n}{b_n} \to 0. \quad (19)$$

Practically, these timescales can be realized by having, e.g., $c_n = \frac{1}{n}, b_n = \frac{1}{n^{0.8}}, a_n = \frac{1}{n^{0.6}}$.

(16), (18) and (17) and form the Two Timescale Stochastic Gradient Algorithm (TTSGA). In a nutshell, TTSGA consists of updating three quantities: LM that determines the waterfilling threshold, the average power consumption and transmission probability. These three quantities are updated on different timescales. In Section VI-A, we show that if each user implements TTSGA then the transmission probability vector converges to a Nash equilibrium. Moreover, the LMs converge to values such that the rate constraints are satisfied.

## VI. ANALYSIS OF TTSGA

In this section, we comment on several properties of TTSGA (17), (18) and (16) such as convergence and fairness. We begin with convergence analysis.

### A. Convergence Analysis

In this section, we prove that:
- The probability iterates under TTSGA converge to a Nash equilibrium under certain conditions.
- Rate constraints are satisfied.

Let $\boldsymbol{\theta}_n = [\theta^1_n, \ldots, \theta^N_n]$ denote the vector of transmission probabilities of the users. For a fixed transmission probability vector $\boldsymbol{\theta} = [\theta^1, \ldots, \theta^N]$, let $\beta^i$ denote the probability of successful transmission for user $i$. It can be expressed as:

$$\beta^i = \theta^i \prod_{j \neq i}(1 - \theta^j). \quad (20)$$

In this case, average power and LM update equations can be expressed as:

$$\mathcal{P}^i_{n+1} = \mathcal{P}^i_n + b_n \theta^i P^i_w(X^i_n), \quad (21)$$

and

$$\lambda^i_{n+1} = \lambda^i_n - a_n(\beta^i U^i_n - \bar{\rho}^i), \ \lambda^i_n \geq 0 \ \forall n. \quad (22)$$

(22) can be expressed as:

$$\lambda^i_{n+1} = \lambda^i_n - a_n(\beta^i U^i_n - \beta^i \bar{U}^i_n + \beta^i \bar{U}^i_n - \bar{\rho}^i), \ \lambda^i_n \geq 0 \ \forall i, \quad (23)$$

where $\bar{U}_n^i$ is the running average of the transmission rate and $\beta^i(U_n^i - \bar{U}_n^i)$ forms a martingale difference sequence. Note that the LM determines the waterfilling threshold and hence the power with which a user transmits in each channel state which in turn determines the average rate that the user achieves. Hence $\bar{U}_n^i$ can be expressed as:

$$\bar{U}_n^i = F^i(\lambda_n^i), \qquad (24)$$

for some continuously differentiable function $F^i(\cdot)$. Using (24), (23) can be expressed as:

$$\lambda_{n+1}^i = \lambda_n^i - a_n(\beta^i U_n^i - \beta^i F^i(\lambda_n^i) + \beta^i F^i(\lambda_n^i) - \bar{\rho}^i), \ \lambda_n^i \geq 0 \ \forall i, \qquad (25)$$

Consider a 'fluid approximation' of (23) with the interpretation that we consider smaller and smaller slot lengths, so that power and transmission probability are interpreted as 'per unit time' quantities instead of 'per slot' quantities. Under this approximation, (22) can be considered to be a noisy discretization of the following ordinary differential equation (o.d.e.):

$$\dot{\lambda}^i(t) = -(\beta^i F^i(\lambda^i(t)) - \bar{\rho}^i). \qquad (26)$$

The set of equilibria of the o.d.e. in (26) $H \triangleq \{\lambda^{i,*} : \beta^i F^i(\lambda^{i,*}) = \bar{\rho}^i\}$. The stability of the o.d.e. (26) allows us to comment on the convergence of the iterates in (22). Using Theorem 2, Chapter 2 (p. 15) of [35] we can claim the following:

*Lemma 1:* For a fixed $\boldsymbol{\theta}$ vector, the LM iterates in (22) converge to the set $H$ of equilibria of (26).

Note that convergence of the LM iterates directly implies that the rate constraint is satisfied with equality.

*Remark 5:* The situation at each user $i$ with fixed $\boldsymbol{\theta}$ has similarities with the single user scenario; the probability of success $\beta^i$ can be treated as an analogue of the transmission probability $\beta$ in the single user scenario. In the multiuser scenario, the probability of success determines the waterfilling threshold $\lambda^i$, higher the probability of success, lower the threshold and hence lower the power consumed. However, the parameter that the user controls is the transmission probability. Increasing the transmission probability increases the probability of success but also increases the probability of collision. The user seeks a balance between these two.

Consider the transmission probability update equation:

$$\theta_{2n+1}^i = \pi_1 \left[ \theta_{2n-1}^i - c_{2n-1} \left( \frac{\mathcal{P}_{2n-1}^i - \mathcal{P}_{2n}^i}{2\delta} \right) \right]. \qquad (27)$$

For the sake of analysis, let us drop the projection operator. We will comment on the projection operation later in this section.

Now, consider a fluid approximation of (27) expressed as the following o.d.e.:

$$\dot{\theta}^i(t) = -\nabla_{\theta^i} \mathcal{P}^i(t), \qquad (28)$$

where $\nabla_{\theta^i}$ denotes the gradient w.r.t. $\theta^i$. The system wide vector o.d.e. can be expressed as:

$$\dot{\boldsymbol{\theta}}(t) = -\nabla_{\boldsymbol{\theta}} \mathbf{P}(t), \qquad (29)$$

Where $\mathbf{P}$ is the power vector $[\mathcal{P}^1, \cdots, \mathcal{P}^N]$. It is clear that $\nabla_{\boldsymbol{\theta}} \mathbf{P}(\cdot)$ acts as a Lyapunov function (See [35] Chapter 10, (10.2.1)), i.e.,

$$\frac{d}{dt}\mathbf{P}(t) = -||\nabla_{\boldsymbol{\theta}} \mathbf{P}(t)||^2 < 0 \qquad (30)$$

Let $G \triangleq \{\boldsymbol{\theta} : \nabla_{\boldsymbol{\theta}} \mathbf{P} = 0\}$ denote the set of local minima of $\mathbf{P}$, i.e., the equilibrium points for this o.d.e.

We now prove that the Hessian $\mathcal{H}$ of $\mathbf{P}$ at $\boldsymbol{\theta} = \boldsymbol{\theta}^*$ under certain conditions is positive definite. This implies that $\boldsymbol{\theta}^* \in G$ denotes a stable equilibrium point of the o.d.e. (29) [36]. In fact, the equilibrium is a Nash equilibrium [36] (See also Proposition 2 of [26]). In order to prove the positive definiteness of $\mathcal{H}$ at $\boldsymbol{\theta} = \boldsymbol{\theta}^*$, we show that it is strictly diagonally dominant which is a sufficient condition for positive definiteness ([37], Section 4.2.1). Next, we evaluate the components of the Hessian $\mathcal{H}$.

Recall that $\mathcal{P}_{\theta^i}^i$ denotes the average power consumed by user $i$ when the transmission probability is $\theta^i$. It consists of two components: $\mathcal{P}_{\beta^i}^i$ which corresponds to the power consumed in successful transmissions and $\mathcal{P}_{\varepsilon^i}^i$ which corresponds to the power wasted in collisions. Let $p(x^i)$ denote that the user $i$ has channel condition $x^i$. The average throughput for user $i$ with a success probability of $\beta^i$ can be expressed as:

$$\beta^i \sum_{x^i} p(x^i) \log_2(1 + \frac{P_{\beta^i}^i(x^i) x^i}{N_0}), \qquad (31)$$

where $P_{\beta^i}^i(x^i)$ is the transmission power in channel state $x^i$ when the success probability is $\beta^i$. From Remark 3, the constraint is satisfied with equality. Hence, using (14) and (31), we can claim that:

$$\beta^i \sum_{x^i} p(x^i) \log_2(1 + \frac{P_{\beta^i}^i(x^i) x^i}{N_0}) = \bar{\rho}^i$$

$$\implies \sum_{x^i} p(x^i) \log_2(1 + \frac{P_{\beta^i}^i(x^i) x^i}{N_0}) = \frac{\bar{\rho}^i}{\beta^i}. \qquad (32)$$

Let $\bar{\rho}^i(x^i)$ denote the rate at which user $i$ transmits in channel state $x^i$ when $\beta^i = 1$. From (32),

$$\log_2(1 + \frac{P_{\beta^i}^i(x^i) x^i}{N_0}) = \frac{\bar{\rho}^i(x^i)}{\beta^i}$$

$$\implies P_{\beta^i}^i(x^i) = \frac{N_0}{x^i}(2^{\frac{\bar{\rho}^i(x^i)}{\beta^i}} - 1). \qquad (33)$$

It can be seen that $P_{\beta^i}^i(x^i)$ is a convex decreasing function of $\beta^i$. Now,

$$\mathcal{P}_{\beta^i}^i = \sum_{x^i} p(x^i) P_{\beta^i}^i(x^i) = \sum_{x^i} p(x^i) \frac{N_0}{x^i}(2^{\frac{\bar{\rho}^i(x^i)}{\beta^i}} - 1). \qquad (34)$$

Hence $\mathcal{P}_{\beta^i}^i$ is also a convex decreasing function of $\beta^i$. $\mathcal{P}_{\theta^i}^i$ can be expressed as:

$$\mathcal{P}_{\theta^i}^i = \frac{\theta^i}{\beta^i} \mathcal{P}_{\beta^i}^i. \qquad (35)$$

*Lemma 2:* The following condition is a sufficient condition for the Hessian $H$ to be positive definite:

$$\frac{1}{\theta^{i,*}} - \sum_{k \neq i} \frac{1}{(1 - \theta^{k,*})} > 0. \qquad (36)$$

*Proof:* Note that,

$$\frac{\partial(\mathcal{P}_{\theta^i}^i)}{\partial \theta^i} = \sum_{x^i} p(x^i)\Big(\frac{N_0 \bar{\rho}^i(x^i)\ln(2)}{(\theta^i)^2(\epsilon^i)^2}\Big)2^{\frac{\bar{\rho}^i(x^i)}{\beta^i}}, \quad (37)$$

where $\epsilon^i = \prod_{j \neq i}(1-\theta^j)$. Moreover for $k \neq i$,

$$\frac{\partial(\mathcal{P}_{\theta^i}^i)}{\partial \theta^k} = \sum_{x^i} p(x^i)\Big(\frac{N_0 2^{\frac{\bar{\rho}^i(x^i)}{\beta^i}}(\epsilon^i \theta^i + \ln(2)\bar{\rho}^i(x^i))}{(1-\theta^k)^2(\epsilon^i)^2 \theta^i x^i}\Big). \quad (38)$$

Furthermore,

$$\frac{\partial^2(\mathcal{P}_{\theta^i}^i)}{\partial(\theta^i)^2} =$$

$$\sum_{x^i} p(x^i)\Big(\frac{N_0 \ln(2)\bar{\rho}^i(x^i) 2^{\frac{\bar{\rho}^i(x^i)}{\beta^i}}(2\epsilon^i \theta^i + \ln(2)\bar{\rho}^i(x^i))}{x^i(\theta^i)^4(\epsilon^i)^3}\Big). \quad (39)$$

Finally,

$$\Big|\frac{\partial^2(\mathcal{P}_{\theta^i}^i)}{\partial(\theta^i \theta^k)}\Big| = \Big|\frac{\partial^2(\mathcal{P}_{\theta^i}^i)}{\partial(\theta^k \theta^i)}\Big| =$$

$$\sum_{x^i} p(x^i)\Big(\frac{N_0 \ln(2)\bar{\rho}^i(x^i) 2^{\frac{\bar{\rho}^i(x^i)}{\beta^i}}(2\epsilon^i \theta^i + \ln(2)\bar{\rho}^i(x^i))}{x^i(\theta^i)^3(\epsilon^i)^3(1-\theta^k)}\Big). \quad (40)$$

From [37], Section 3.4.10, the following two conditions are sufficient for strict diagonal dominance of $H$ at $\boldsymbol{\theta} = \boldsymbol{\theta}^*$; first: $|\frac{\partial^2(\mathcal{P}_{\theta^i}^i)}{\partial(\theta^i)^2}| > 0 \ \forall i$ which is easily verified from (39); and second:

$$\Bigg[\Big|\frac{\partial^2(\mathcal{P}_{\theta^i}^i)}{\partial(\theta^i)^2}\Big| - \sum_{k \neq i}\Big|\frac{\partial^2(\mathcal{P}_{\theta^i}^i)}{\partial(\theta^i \theta^k)}\Big|\Bigg]_{\boldsymbol{\theta}=\boldsymbol{\theta}^*} > 0. \quad (41)$$

With little algebraic manipulation on (39) and (40), it can be shown that this condition is equivalent to:

$$\frac{1}{\theta^{i,*}} - \sum_{k \neq i}\frac{1}{(1-\theta^{k,*})} > 0. \quad (42)$$

∎

Positive definiteness of Hessian enables us to claim the following:

*Theorem 1:* If there is a stable equilibrium point $\boldsymbol{\theta}^* \in G$, such that (42) is satisfied, then for any initial transmission probabilities $\boldsymbol{\theta}_0$ the dynamics $\boldsymbol{\theta}(t)$ in (29) converge to $\boldsymbol{\theta}^*$ asymptotically.

Note that so far, we have studied the convergence of the o.d.e. (29). However, we are really concerned about the convergence of the primal TTSGA iterates in (27). Theorem 2, Chapter 2 (p. 15) of [35] allows us to claim the convergence of these iterates:

*Theorem 2:* If there is a stable equilibrium point $\boldsymbol{\theta}^* \in G$ of the o.d.e. (29), then for any initial transmission probabilities $\boldsymbol{\theta}_0$ the $\boldsymbol{\theta}_n$ iterates in (27) converge to $\boldsymbol{\theta}^*$.

Finally, using Theorem 2, Chapter 6 (p. 66) of [35] we claim that the coupled iterates converge, i.e.,

*Theorem 3:* The coupled iterates $(\lambda^i, \theta^i)$ in (16) and (27) converge to their respective equilibrium values.

Note that since the $\theta^i$ iterates are updated on the slower timescale, these iterates see converged values of LMs at each update instant. Finally, when the $\boldsymbol{\theta}$ iterates converge to the equilibrium, the corresponding LM values while ensuring that rate constraints are satisfied, also determine the correct long term power consumption.

*Remark 6:* One way to ensure the condition in (42) is to enforce a limit on the maximum probability with which a user can transmit in a slot. We already have this mechanism in place through the limit $\omega$ on transmission probability. The second term in (42) takes its maximum value when $\theta_*^k = \omega, \ \forall k$, while the first term takes its minimum value when $\theta^{i,*} = \omega$. For the minimum value of the LHS to be greater than 0 we require that:

$$\frac{1}{\omega} - \frac{N-1}{1-\omega} > 0 \implies \omega < \frac{1}{N}. \quad (43)$$

Condition (43) forces the transmission probability iterates in the interval $[0, \frac{1}{N})$. If there exists an equilibrium point in the interval $[0, \frac{1}{N})$, the iterates converge to such an equilibrium.

### B. Equilibrium as a Best Response

Nash equilibrium embodies the notion of best response offered by a player to the strategies of the other players. Unilateral deviation from the equilibrium does not result in an increase in the utility for any player. In the present case also the transmission probability is a best response to the transmission probabilities of the other players. Unilateral deviation from the equilibrium transmission probability does not result in a decrease in the average power consumption for a player. This is because decreasing the transmission probability by, say, user $i$ from its equilibrium value decreases both - its success probability $\beta^i$ and collision probability $\varepsilon^i$. This increases $\mathcal{P}_{\beta^i}^i$ and reduces $\mathcal{P}_{\varepsilon^i}^i$; the net effect being that there is an increase in the overall power consumption. On the other hand, increasing the transmission probability from its equilibrium value increases $\beta^i$ but also increases $\varepsilon^i$. This reduces $\mathcal{P}_{\beta^i}^i$ but increases $\mathcal{P}_{\varepsilon^i}^i$; the net effect being that there is again an increase in the overall power consumption.

### C. Multiuser Penalty

Let $\mathcal{P}_1^i \ (\beta^i = 1)$ denote the average consumed by a user $i$ in the single user scenario with successful transmission in every slot. In the multiuser scenario, $\beta^i < 1$. Since the user is able to successfully communicate only during a fraction of the slots, the rate constraint appears to be appropriately scaled, resulting in a corresponding scaling of the LM or the waterfilling threshold. This results in a corresponding increase in the power consumption. This is the penalty that the user pays in operating in a multiuser environment. We term this penalty as the *multiuser penalty*. Note that larger the number of users, potentially larger is the penalty paid by the user for obtaining a certain throughput.

### D. Fairness

A user does not have an incentive for arbitrarily increasing the transmission probability, it can be increased only when the reduction in power consumption due to increase in success probability outweighs the corresponding wastage due to increased probability of collision. The useful power consumption depends on two factors: the rate constraint and average channel condition. A user having a high rate constraint would require high rates in each channel state as compared to a user having a lower rate constraint but with same channel statistics. A user with higher rate constraint would therefore require more 'useful' power. Increasing the transmission rate increases the success probability and reduces the 'useful' power requirement. Therefore, such a user would have a higher transmission probability but this also results in higher collision rate and higher 'wasted' power. Similar arguments can be made for users with same rate requirement and different average channel conditions. This discussion implies that users having higher rate requirement or poorer channel consume more power thus ensuring fairness.

In the next section, we simulate TTSGA in a discrete event simulator. Our objective is to demonstrate that TTSGA satisfies the rate constraints through simulation studies.

## VII. Experimental Evaluation

In this section, we simulate a single cell wireless system where $N$ users communicate with the base station on the uplink. The user applications require average rate guarantees. Packets are generated at the application layer and are possibly of variable sizes. At the MAC layer, we assume that each MAC fragment is of constant length equal to $\ell = 2000$ bits. We assume that the system has a bandwidth $W$ of 10 MHz. Each user transmits at a constant power of 1 Watt. We assume $\omega = 0.1$. We simulate a Rayleigh channel for each user. For a Rayleigh model, channel state $X^i$ is an exponentially distributed random variable with mean $\alpha^i$ and probability density function expressed as $f_X(x) = \frac{1}{\alpha^2}\exp\left(\frac{-x^2}{2\alpha^2}\right)$, $x \geq 0$. We discretize the channel into eight equal probability bins, with the boundaries specified by { $(-\infty, -8.47$ dB), $[-8.47$ dB, $-5.41$ dB), $[-5.41$ dB, $-3.28$ dB), $[-3.28$ dB, $-1.59$ dB), $[-1.59$ dB, $-0.08$ dB), $[-0.08$ dB, $1.42$ dB), $[1.42$ dB, $3.18$ dB), $[3.18$ dB, $\infty$ ) }. For each bin, we associate a channel state and the state space $\mathcal{X}$ = { $-13$ dB, $-8.47$ dB, $-5.41$ dB, $-3.28$ dB, $-1.59$ dB, $-0.08$ dB, $1.42$ dB, $3.18$ dB}. This discretization of the state space of $X^i$ has been justified in [38]. For the sake of simplicity, we assume that the rate of transmission for user $i$ in slot $n$ can be determined using the following capacity relation:

$$U_n^i = W \times \log_2(1 + \frac{P_w X_n^i}{N_0 W}). \tag{44}$$

We consider a system with $N = 20$ user destination pairs. We divide the user destination pairs into 2 groups (Group 1 and Group 2) of 10 pairs each. We consider two scenarios: rate variation and channel variation. We present the results after averaging over 20 simulation runs each consisting of simulating the algorithm for $100,000$ slots.

*Scenario 1: Rate Variation:* In this scenario, we vary the average rate constraints for the pairs in Group 2 in successive experiments while keeping the average rate constraints for the pairs in Group 1 constant in all the experiments. For Group 2, the rate constraints are varied as 50-110 Kbps in steps of 10 Kbps in successive experiments, while the rate constraints for Group 1 are kept at 50 Kbps in all experiments. The mean channel state $\alpha$ for all the destinations is kept at $-3.28$ dB $(0.4698)$ for all the experiments. In each slot, we generate the channel state using the exponential distribution with mean $\alpha$ and subsequently discretize it using the probability bins as mentioned above. Each user makes the scheduling decision using its transmission probability $\theta_n^i$ in each slot. Based on the feedback received from the base station, it then determines the new value of the transmission probability. We select two users, at random from Group 1 and Group 2. For these pairs, we determine the power consumed, the stable transmission probabilities and rate achieved within each experiment and plot these in Figures 3, 4 and 5 respectively. It can be seen from Figure 5 that the rate constraint satisfied. Moreover, from Figures 3, 4 it can be seen that as the rate constraint is increased, the power expended and the transmission probability increase.

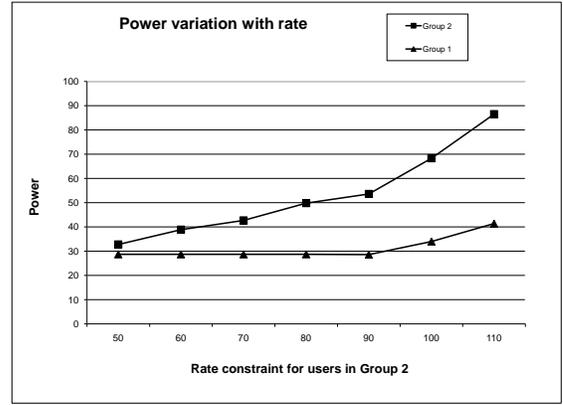

Fig. 3. Power expended for various average rate constraints

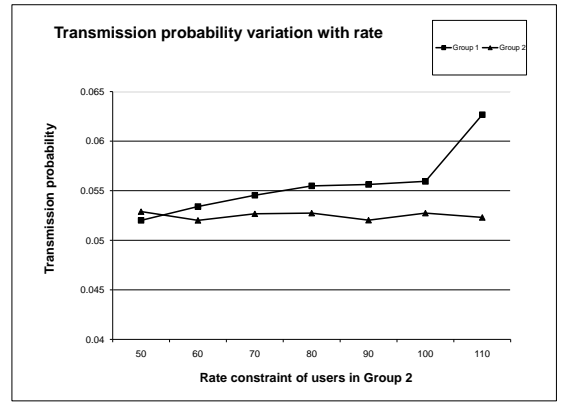

Fig. 4. Transmission probability (TP) for various average rate constraints

*Scenario 2: Channel Variation:* In this scenario, we vary the average channel state for the pairs in Group 2 in successive experiments while keeping the average channel state for the pairs in Group 1 constant in all experiments. For the pairs in Group 2, the average channel state $\alpha$ is varied as $0.05$ ($-13$ dB), $0.1422$ ($-8.47$ dB), $0.2877$ ($-5.41$ dB), $0.4698$ ($-3.28$ dB), $0.6934$ ($-1.59$ dB), $0.9817$ ($-0.08$ dB), $1.3867$ ($1.42$ dB) in successive experiments, while the average channel state for the pairs in Group 1 is kept at $0.4698$ ($-3.28$ dB) in all experiments. The average rate constraint is kept constant at 50 Kbps for all the pairs for all experiments. In each time slot, we generate the channel state using the exponential distribution with mean $\alpha$ and subsequently discretize it using the probability bins as mentioned above. We select two users at random from Group 1 and Group 2. For these pairs, we determine the power consumer, the stable transmission probabilities and rate achieved within each experiment and plot these in Figures 6, 7 and 8 respectively. 8 demonstrates that the rate constraints are satisfied. Moreover, it can be seen from Figures 6, 7 that as the average channel state improves, the transmission probability reduces thus reducing the average power expenditure.

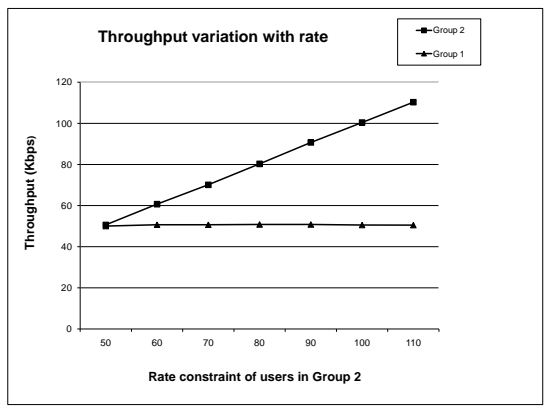

Fig. 5. Throughput for various average rate constraints

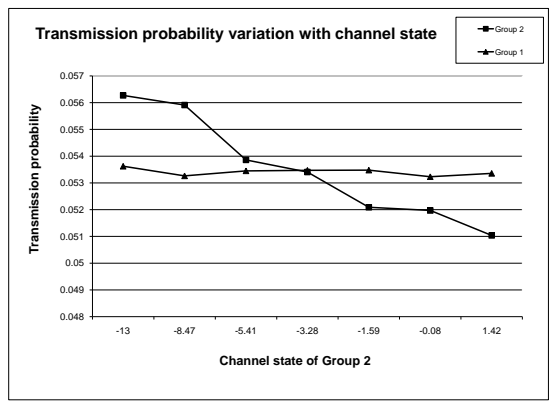

Fig. 7. Transmission probability (TP) for various average channel states

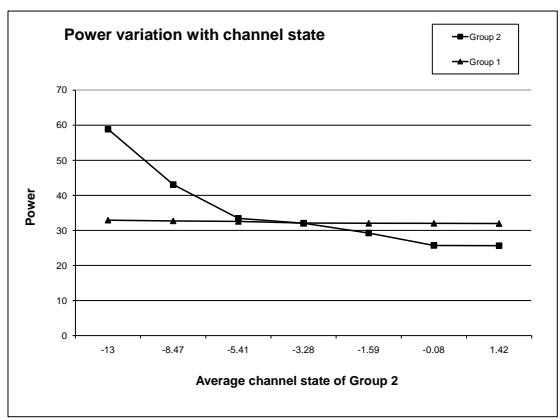

Fig. 6. Power expended for various average channel states

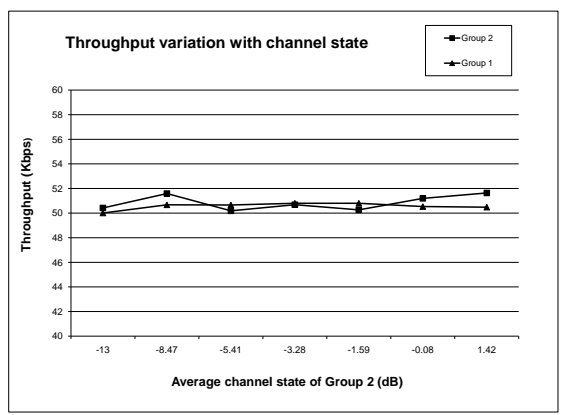

Fig. 8. Throughput for various average channel states

## VIII. PRACTICAL IMPLEMENTATION

In this section, we first describe in brief the operation of Carrier Sense Multiple Access/Collision Avoidance (CSMA/CA) access control protocol employed in IEEE 802.11 [39]. We then describe a protocol for practically implementing TTSGA. This protocol is based on a modification of IEEE 802.11 CSMA/CA access framework. We then discuss implementation details relevant to TTSGA within this framework.

### A. IEEE 802.11 CSMA/CA Protocol

In CSMA, a user that desires to transmit has to first sense the channel for a predetermined amount of time. If the channel is sensed 'idle' then the user proceeds with its transmission. If the channel is sensed 'busy' then the user has to defer its transmission for a random duration of time (backoff). This is done to reduce the collision probability.

Wireless local area networks (LANs) have the 'hidden terminal' problem, wherein a transmitter $T_1$ transmitting to a receiver $R$ that is not in transmission range of another transmitter $T_2$ does not know whether $T_2$ is transmitting to $R$. This can lead to collisions at $R$. This hidden terminal problem is avoided through the Collision Avoidance (CA) mechanism of IEEE 802.11. The CA mechanism involves transmission of Request to Send (RTS) packet by transmitter $T_1$ to receiver $R$ (refer to Figure 9). The receiver, if idle, sends the Clear to Send (CTS) packet back to $T_1$. This alerts all the transmitters in the range of $T_1$ and $R$ of the data transfer between $T_1$ and $R$. The other transmitters then suspend their transmissions till the duration of the data transfer (Data portion in Figure 9).

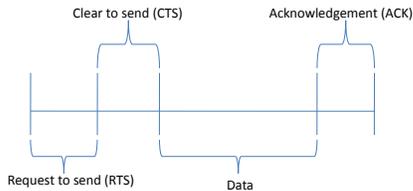

Fig. 9. Slot structure

Transmission of the Acknowledgement (ACK) packet from $R$ to $T_1$ signals successful transfer.

If $T_1$ after transmitting RTS does not receive CTS for a certain period of time, it infers that it was involved in a collision. In order to avoid further collisions, it backs off or defers its transmission using an 'exponential backoff' mechanism prescribed by the protocol. This mechanism consists of backing off randomly by selecting a backoff duration from a window $[0, W]$ referred to as the contention window. Each time a transmitter is involved in a collision, it doubles its contention window and chooses a random backoff duration from that window.

### B. A Protocol for Practical Implementation of TTSGA

We now describe a protocol for practically implementing TTSGA. This protocol replaces or supplements the exponential backoff mechanism of the IEEE CSMA/CA protocol. A user first contends for the channel by transmitting a RTS packet based on the current transmission probability computed by TTSGA. The RTS packet is sent at the most robust rate to guard against channel fading. CSI can be estimated by the users based on a pilot transmitted by the access point. In a TDD system, the same CSI can be used by the users for uplink transmissions. CSI can be conveyed to the base station using ranging messages. There might be multiple users contending for the channel by transmitting RTS packets to the access point. If there is no collision between these RTS packets, the access point responds with a CTS packet. This packet also contains information regarding rate at which the user should transmit based on the CSI. If the user receives the CTS packet, it utilizes the transmission rate information in the CTS packet and transmits accordingly. If the access point receives the data transmitted by the user devoid of errors, it sends an ACK packet back to the user. The user then enters the contention mode again by transmitting the RTS packet with a recomputed transmission probability based on TTSGA.

When the user receives a CTS packet corresponding to a different user, it suspends its transmission till it receives an ACK packet corresponding to that data transfer. This signals the completion of the transmission between the base station and the associated user. Reception of the appropriate CTS packet allows the user to determine whether it was involved in a collision or not. If an appropriate CTS packet is received, the user determines the number of packets that can be transmitted in the Data part of the slot and transmits those packets. It then recomputes the transmission probability based on whether the transmission was successful or not. Thereafter, the user contends for the channel by transmitting a RTS packet with the newly computed transmission probability.

### C. Practical Implementation of TTSGA

Based on online primal-dual computations in (16), (18) and (17), and the protocol suggested in previous sub-section, user $i$ implements the access control scheme. The user is aware of the value of channel state $X^i$ in each slot through the CTS packet. The number of packets to be transmitted is then determined using (8). The user $i$ transmits with probability $\theta^i + \delta$ in odd numbered attempts and with probability $\theta^i - \delta$ in even numbered attempts. The transmission probability is adjusted in odd numbered attempts. If a transmission is successful, $U^i$ packets are received at the base station and an acknowledgement packet is received and the LM $\lambda^i$ is appropriately updated. The algorithm thus continues. The complete scheme is explained in Algorithm 1.

---

1: Initialize the LM $\lambda_0^i \leftarrow 0$, $\theta_0^i \leftarrow \theta_0$, $n \leftarrow 1$, channel state $X_0^i \leftarrow 0$.
2: **while** TRUE **do**
3:     Transmit RTS packet with probability $\theta_n^i$.
4:     **if** CTS received **then**
5:        Use CSI $X_n^i$ in CTS packet to determine $U_n^i$.
6:        Transmit $U_n^i$ packets.
7:     **else**
8:        $U_n^i \leftarrow 0$.
9:     **end if**
10:    **if** ACK received **then**
11:       $J_n^i \leftarrow 1$.
12:    **else**
13:       $J_n^i \leftarrow 0$.
14:    **end if**
15:    Update the LM $\lambda_n^i$ using (16).
16:    Update transmission probability $\theta_n^i$ using (17).
17:    $n \leftarrow n + 1$.
18: **end while**

**Algorithm 1:** Two Timescale Stochastic Gradient Algorithm (TTSGA)

---

*Remark 7:* Note that in this paper, we assume that the users transmit at a 'reliable' rate. Under this assumption, there are no transmission errors and users would always receive an ACK packet. The purpose of including ACK in the protocol is to indicate to the other users about the end of a transmission so that they can start contending for channel access. The case where transmission errors do occur and ACK reception (non-reception) is an indication of successful (unsuccessful) transmission forms an interesting future work.

## IX. CONCLUSIONS

In this paper, we have considered uplink transmissions in a single cell multiuser wireless system. The base station does not coordinate the transmission of the users, hence the users employ random access communication. In each slot, the users obtain a $(0, 1, e)$ feedback from the base station. The users have an objective of minimizing their long term power consumption while achieving certain long term average rates. We have modeled the situation as a constrained repeated non-cooperative game where users have knowledge of their utility function only. We have proposed a two timescale stochastic gradient algorithm (TTSGA) at the users in order to tune their transmission probabilities. The algorithm includes a 'waterfilling threshold update mechanism' that appropriately tunes the waterfilling threshold for each user and ensures that the rate constraints are satisfied. We have proved that under the algorithm the transmission strategies converge to a Nash equilibrium and that the rate constraints are satisfied. Moreover, our simulation studies have also demonstrated that the rate constraints are satisfied.